\documentclass[12pt]{article}

\usepackage{bm}

\begin{document}

\title{The Feynman-Dyson Propagators for Neutral Particles (Locality or Non-locality?)\thanks{Presented at the XII Taller of DGyFM SMF, Nov. 27 - Dec. 1, 2017. Guadalajara, M\'exico}\\}

\author{{\bf Valeriy V. Dvoeglazov}\\UAF, Universidad Aut\'onoma de Zacatecas, M\'exico\\E-mail: valeri@fisica.uaz.edu.mx}

\date{\empty}

\maketitle

\begin{abstract}An analog of the $S=1/2$ Feynman-Dyson propagator  is presented in the framework of the $S=1$ Weinberg's theory.
The basis for this construction is the concept of the Weinberg field as a system of  four field functions differing  by parity  and  by dual transformations.

Next, we analyze the recent controversy in the definitions of the Feynman-Dyson propagator for the field operator containing the $S=1/2$ self/anti-self charge conjugate states in the papers by D. Ahluwalia et al. and by W. Rodrigues Jr. et al. The solution of this mathematical controversy is obvious. It is related to the necessary doubling of  the Fock Space (as in the Barut and Ziino works), thus extending the corresponding Clifford Algebra. However, the logical interrelations of different mathematical foundations with the physical interpretations are not so obvious. Physics should choose only one correct formalism –- it is not clear, why two correct mathematical formalisms  (which are based on the same postulates) lead to different physical results?
\end{abstract}



\large{

\section{The Weinberg Propagators.}

Accordingly  to the Feynman-Dyson-Stueckelberg ideas,
a causal propagator  has to be constructed
by using  the formula (e.~g., Ref.~\cite[p.91]{Itzykson})
\begin{eqnarray}
\lefteqn{\hspace{-6mm}S_F (x_2, x_1) =\sum_\sigma\int \frac{d^3 p}{(2\pi)^3}\frac{m}{E_p} \left [
\theta (t_2 -t_1) \, a \,\,
u^\sigma (p)  \overline u^{\sigma} (p) e^{-ipx} + \right. }\nonumber\\
&&\left. \qquad\qquad  +\, \,\theta (t_1 - t_2) \, b \,\,
v^\sigma (p)  \overline v^\sigma (p) e^{ipx} \right ] \, ,
\end{eqnarray}
$x=x_2 -x_1$. In the $S=1/2$ Dirac theory  it results  to
\begin{equation}\label{dp}
S_F (x) = \int \frac{d^4 p}{(2\pi)^4} e^{-ipx} \frac{\hat p +m}{p^2 -m^2
+i\epsilon} \,,
\end{equation}
provided that the constant $a$ and $b$ are determined by imposing
\begin{equation}
(i\hat \partial_2 -m) S_F (x_2, x_1) =\delta^{(4)} (x_2 -x_1) \, ,
\end{equation}
namely, $a=-b=1/ i$ .

However,  attempts to construct the covariant propagator in this way
have failed in the framework of the Weinberg theory,
Ref.~\cite{Weinberg}, which is a generalization of the Dirac ideas to
higher spins. For instance, on the page B1324 of Ref.~\cite{Weinberg}\,
Weinberg writes:

\smallskip

``{\it Unfortunately,
the propagator arising from Wick's theorem is  NOT equal to the covariant
propagator
except for $S=0$ and $S=1/2$. The trouble is that the derivatives act on the
$\epsilon (x) = \theta (x) - \theta (-x)$ in $\Delta^C (x)$ as well as on
the functions\footnote{In the cited paper $\Delta_1(x) \equiv i \left [\Delta_+
(x) + \Delta_+ (-x)\right ]$ and $\Delta (x) \equiv \Delta_+ (x) - \Delta_+
(-x)$ have been used.
 $i\Delta_+ (x) \equiv \frac{1}{(2\pi)^3} \int \frac{d^3 p}{2E_p} \exp
(ipx)$ is the particle Green function.} $\Delta$ and $\Delta_1$. This
gives rise to extra terms proportional to equal-time $\delta$ functions
and their derivatives\ldots The cure is well known: \ldots compute the
vertex factors using only the original covariant part of the Hamiltonian
${\cal H}$; do not use the Wick propagator for internal lines; instead use
the covariant propagator.}

\smallskip

The propagator proposed  in Ref.~\cite{Ah-pro} is the causal propagator.
However, the old problem remains: the Feynman-Dyson propagator
is not the Green function of the Weinberg equation. As mentioned,
the covariant propagator proposed by Weinberg propagates kinematically
spurious solutions~\cite{Ah-pro}.

The aim of my paper is to consider
the problem of constructing the propagator in
the framework of the model given in~\cite{D1}.
The concept of
the Weinberg field `doubles'  has been proposed there.
It is based on the equivalence between
the Weinberg field and the  antisymmetric tensor field, 
which can be described by both $F_{\mu\nu}$ and
its dual $\tilde F_{\mu\nu}$.
These field functions may be used to form a parity doublet. An essential
ingredient of my consideration
is the idea of combining the Lorentz and the dual
transformation. 

The set of four equations has been proposed in Ref.~\cite{D1}.
For the functions $\psi_1^{(1)}$ and $\psi_2^{(1)}$, connected with the
first one by  the dual (chiral, $\gamma_5$)  transformation,
the equations are
\begin{eqnarray}\label{eq:a1}
(\gamma_{\mu\nu} p_\mu p_\nu +m^2 )\psi_1^{(1)} &=& 0\,,\\ \label{eq:a2}
(\gamma_{\mu\nu} p_\mu p_\nu - m^2) \psi_2^{(1)} &=& 0 \,.
\end{eqnarray}
For the field functions
connected with $\psi_1^{(1)}$ and $\psi_2^{(1)}$ by $\gamma_5\gamma_{44}$
transformations the set of equations is written:
\begin{eqnarray}\label{eq:a11}
\left [\widetilde \gamma_{\mu\nu}p_\mu p_\nu - m^2\right ] \psi_1^{(2)} &=&0
\,,\\
\label{eq:a21}
\left [\widetilde \gamma_{\mu\nu} p_\mu p_\nu + m^2 \right ] \psi_2^{(2)} &=&0
\,,
\end{eqnarray}
where $\widetilde \gamma_{\mu\nu} = \gamma_{44} \gamma_{\mu\nu} \gamma_{44}$
is connected with the Barut-Muzinich-Williams $S=1$ matrices~\cite{Barut, Hammer}.

In the cited paper  I  have used the plane-wave expansion:
\begin{eqnarray}\label{pl1}
\psi_1 (x) &=&\sum_\sigma \int \frac{d^3 p}{(2\pi)^3} \frac{1}{m \sqrt{2E_p}}
\left [ u_1^\sigma (\vec p) a_\sigma (\vec p) e^{ipx} +v_1^\sigma (\vec p)
b^\dagger_\sigma (\vec p) e^{-ipx} \right ]\,,\nonumber\\
&&\label{pl2}\\
\psi_2 (x) &=&\sum_\sigma \int \frac{d^3 p}{(2\pi)^3} \frac{1}{m\sqrt{2E_p}}
\left [ u_2^\sigma (\vec p) c_\sigma (\vec p) e^{ipx} +v_2^\sigma (\vec p)
d^\dagger_\sigma (\vec p) e^{-ipx} \right ]\,,\nonumber\\
&&
\end{eqnarray}
where $E_p=\sqrt{\vec p^{\,2} +m^2}$,  in order
to prove that  one can describe a
$S=1$ quantum particle with transversal components in the framework
of  the Weinberg and/or  the antisymmetric tensor theory.

The corresponding `bispinors'
in the momentum space coincide with the Tucker-Hammer ones within
a normalization.\footnote{They  also coincide with the Ahluwalia
{\it et al.} ones within a unitary transformation.} Their
explicit forms are
\begin{eqnarray}\label{b1}
u_1^{\sigma\, (1)} (\vec p)= v_1^{\sigma\, (1)}
(\vec p) =\frac{1}{\sqrt{2}}\pmatrix{\left [m+ (\vec S\cdot\vec p)
+{(\vec S \cdot\vec p)^2 \over  (E+m)}\right ]\xi_\sigma \cr \left [ m  -
(\vec S\cdot\vec p) +{(\vec S \cdot\vec p)^2 \over  (E+m)}\right ]
\xi_\sigma} \,,
\end{eqnarray}
and
\begin{eqnarray}\label{b2}
u_2^{\sigma\,(1)} (\vec p)= v_2^{\sigma\,(1)} (\vec p)
=\frac{1}{\sqrt{2}}\pmatrix{\left [m+
(\vec S\cdot\vec p) +{(\vec S \cdot\vec p)^2 \over  (E+m)}\right ]\xi_\sigma \cr
\left [ - m  + (\vec S\cdot\vec p) - {(\vec S\cdot \vec p)^2
\over  (E+ m)}\right ] \xi_\sigma}\,.
\end{eqnarray}
Thus,  $u_2^{(1)} (\vec p) = \gamma_5 u_1^{(1)} (\vec p)$ and
$\overline u_2^{(1)} (\vec p) = -\overline u_1^{(1)} (\vec p)\gamma_5$.

The bispinors
\begin{eqnarray}\label{b11}
u_1^{\sigma\, (2)} (\vec p)= v_1^{\sigma\, (2)}
(\vec p) =\frac{1}{\sqrt{2}}\pmatrix{\left [m- (\vec S\cdot\vec p)
+{(\vec S\cdot \vec p)^2 \over  (E+m)}\right ]\xi_\sigma \cr \left [ -m  -
(\vec S\cdot\vec p) -{(\vec S\cdot \vec p)^2 \over  (E+m)}\right ]
\xi_\sigma}\,,
\end{eqnarray}
\begin{eqnarray}\label{b21}
u_2^{\sigma\,(2)} (\vec p)= v_2^{\sigma\,(2)} (\vec p)
=\frac{1}{\sqrt{2}}\pmatrix{\left [-m+
(\vec S\cdot\vec p) -{(\vec S\cdot \vec p)^2 \over  (E+m)}\right ]\xi_\sigma \cr
\left [  -m  - (\vec S\cdot\vec p) - {(\vec S\cdot \vec p)^2
\over  (E+ m)}\right ] \xi_\sigma}
\end{eqnarray}
satisfy Eqs. (\ref{eq:a11}) and (\ref{eq:a21}) written in the momentum space.
Thus,
$u_1^{(2)} (\vec p) = \gamma_5\gamma_{44} u_1^{(1)} (\vec p)$,
$\overline u_1^{(2)} = \overline u_1^{(1)} \gamma_5\gamma_{44}$,
$u_2^{(2)} (\vec p) = \gamma_5\gamma_{44} \gamma_5 u_1^{(1)} (\vec p)$
and $\overline u_2^{(2)} (\vec p) = - \overline u_1^{(1)}\gamma_{44}$.

Let me check, if the sum of four equations
\begin{eqnarray}
&&\hspace*{-1cm}\left [ \gamma_{\mu\nu} \partial_\mu \partial_\nu -m^2 \right ]
 \int  \frac{d^3 p}{(2\pi)^3 2E_p}
\left [ \theta (t_2 -t_1) \, a\,\,\,   u_1^{\sigma\,(1)} (p)   \overline
u_1^{\sigma\,(1)} (p) e^{ipx}+\right .\nonumber\\
&&\left.  \qquad\qquad+\theta (t_1 -t_2) \, b \,\,\, v_1^{\sigma\,(1)} (p)
  \overline  v_1^{\sigma\,(1)} (p) e^{-ipx} \right  ] +\nonumber\\
&+& \left [ \gamma_{\mu\nu} \partial_\mu \partial_\nu + m^2 \right  ]  \int
\frac{d^3 p}{(2\pi)^3 2E_p}
\left [ \theta (t_2 -t_1) \, a\,\,\,   u_2^{\sigma\,(1)} (p)   \overline
u_2^{\sigma\,(1)} (p) e^{ipx}+
\right. \nonumber\\
&&\left. \qquad\qquad+\theta (t_1 -t_2) \, b \,\,\, v_2^{\sigma\,(1)} (p)
  \overline  v_2^{\sigma\,(1)} (p) e^{-ipx}\right  ] +\nonumber\\
&+&\left [ \widetilde \gamma_{\mu\nu} \partial_\mu \partial_\nu + m^2 \right  ]
\int \frac{d^3 p}{(2\pi)^3 2E_p}
\left [ \theta (t_2 -t_1) \, a\,\,\,  u_1^{\sigma\,(2)} (p)   \overline
u_1^{\sigma\,(2)} (p) e^{ipx}+ \right.\nonumber\\
&&\left. \qquad\qquad+\theta (t_1 -t_2) \, b \,\,\, v_1^{\sigma\,(2)} (p)
  \overline  v_1^{\sigma\,(2)} (p)e^{-ipx} \right ] +\\
&+&\left [\widetilde \gamma_{\mu\nu} \partial_\mu \partial_\nu - m^2 \right  ] \int
\frac{d^3 p}{(2\pi)^3 2E_p}
\left [ \theta (t_2 -t_1) \, a\,\,\,   u_2^{\sigma\,(2)} (p)   \overline
u_2^{\sigma\,(2)} (p)  e^{ipx} +\right.\nonumber\\
&&\left.  \qquad\qquad+\theta (t_1 -t_2) \, b \,\,\, v_2^{\sigma\,(2)} (p)
  \overline  v_2^{\sigma\,(2)} (p) e^{-ipx} \right ] =
\delta^{(4)} (x_2 -x_1)\nonumber
\end{eqnarray}
can be satisfied by the definite choice of $a$ and $b$.
The relation  $ u_i (p) =  v_i (p)$ for bispinors in the momentum space
had been used in Ref.~\cite{D1}.  In the process of
calculations  I  assume
that the 3-'spinors' are normalized to $\delta_{\sigma\sigma^\prime}$\, .

The simple calculations give
\begin{eqnarray}
\lefteqn{\partial_\mu \partial_\nu  \left [ a\, \theta (t_2 -t_1)\, e^{ip(x_2
-x_1)} + b\, \theta (t_1 -t_2)\, e^{-ip(x_2 -x_1)} \right ]=}\\
&=& - \left [ a\, p_\mu p_\nu \theta (t_2 - t_1)
\exp \left [ ip(x_2 -x_1)\right ] +
b\,  p_\mu p_\nu  \theta (t_1 -t_2)
\exp \left [ -ip (x_2 -x_1) \right ] \right
] + \nonumber\\
&+& a\left [ - \delta_{\mu 4} \delta_{\nu 4} \delta^{\,\,\prime}
(t_2 -t_1) +i (p_\mu \delta_{\nu 4} +p_\nu \delta_{\mu 4}) \delta (t_2 -t_1)
\right ] \exp \left [i \vec p
(\vec x_2 - \vec x_1)\right ] +\nonumber\\
&+& b\, \left [ \delta_{\mu 4} \delta_{\nu 4} \delta^{\,\,\prime}
(t_2 -t_1) + i (p_\mu \delta_{\nu 4} +p_\nu \delta_{\mu 4})
\delta (t_2 -t_1) \right ] \exp \left [-i\vec p
(\vec x_2 - \vec x_1)\right ]\,;\nonumber
\end{eqnarray}
and
\begin{eqnarray}
u_1^{(1)}\overline u_1^{(1)} ={1\over 2} \pmatrix{m^2 & S_p \otimes S_p\cr
\overline S_p \otimes \overline S_p &m^2\cr},
u_2^{(1)}\overline u_2^{(1)} = {1\over 2}\pmatrix{-m^2 & S_p \otimes S_p\cr
\overline S_p \otimes \overline S_p &-m^2\cr},\nonumber\\
\hspace{-10mm}
\end{eqnarray}
\begin{eqnarray}
u_1^{(2)}\overline u_1^{(2)} ={1\over 2} \pmatrix{-m^2 &
\overline S_p \otimes \overline
 S_p\cr S_p \otimes  S_p &-m^2\cr},
 u_2^{(2)}\overline u_2^{(2)} = {1\over 2}
\pmatrix{m^2 & \overline S_p \otimes \overline S_p\cr S_p
\otimes  S_p &m^2\cr},\nonumber\\
\hspace{-10mm}
\end{eqnarray}
where
\begin{eqnarray}
S_p &=& m + (\vec S \cdot\vec p) +\frac{(\vec S \cdot\vec p)^2}{E+m}\,,\\
\overline S_p &=& m - (\vec S \cdot \vec p) + \frac{(\vec S \cdot \vec p)^2}{E+m}\,.
\end{eqnarray}
Due to
\begin{eqnarray}\left [E_p - (\vec S\cdot \vec p)\right ]  S_p \otimes S_p &=& m^2 \left [ E_p
+ (\vec S\cdot \vec p)\right ]\,,\\
\left [E_p + (\vec S\cdot \vec p)\right ] \overline S_p
\otimes \overline S_p &=& m^2 \left [ E_p - (\vec S\cdot \vec p)\right ]\,.
\end{eqnarray}
one  can conclude: the  generalization of
the notion of causal  propagators is  admitted by using the
`Wick's formula' for the time-ordered particle operators
provided that  $a=b=1/ 4im^2$. It is necessary to  consider
all four equations, Eqs. (\ref{eq:a1})-(\ref{eq:a21}). Obviously, this is related to the 12-component formalism, which I presented 
in~\cite{D1}.

The $S=1$ analogues of the formula (\ref{dp})  for the Weinberg propagators
follow immediately. In the Euclidean metrics they are:
\begin{equation}\label{propa1}
S_F^{(1)} ( p ) \sim -\frac{1}{i(2\pi)^4  (p^2  +m^2 -i\epsilon)} \left [
\gamma_{\mu\nu} p_\mu p_\nu   -  m^2  \right ]\,,
\end{equation}
\begin{equation}\label{propa2}
S_F^{(2)} ( p ) \sim -\frac{1}{i(2\pi)^4  (p^2  +m^2 -i\epsilon)} \left [
\gamma_{\mu\nu} p_\mu p_\nu   +  m^2  \right ]\,,
\end{equation}
\begin{equation}\label{propa3}
S_F^{(3)} ( p ) \sim -\frac{1}{i(2\pi)^4  (p^2  +m^2 -i\epsilon)} \left [
\widetilde\gamma_{\mu\nu} p_\mu p_\nu   +  m^2  \right ] \,,
\end{equation}
\begin{equation}\label{propa4}
S_F^{(4)} ( p ) \sim -\frac{1}{i(2\pi)^4  (p^2  +m^2 -i\epsilon)} \left [
\widetilde \gamma_{\mu\nu} p_\mu p_\nu   -  m^2  \right ] \,.
\end{equation}

We should use the obtained set of  Weinberg propagators
(\ref{propa1},\ref{propa2},\ref{propa3},\ref{propa4})
in the perturbation calculus of scattering amplitudes.
In Ref.~\cite{Dvoegl-IJTP} the amplitude for the interaction
of two $2(2S+1)$ bosons has been obtained on the basis
of the use of one field only and it is obviously incomplete,
see also Ref.~\cite{Hammer}. But, it is interesting to note
that the spin structure was proved there
to be the same,  regardless we consider
the two-Dirac-fermion interaction or the two-Weinberg($S=1$)-boson
interaction. However, the denominator slightly differs ($1/\vec \Delta^2
\rightarrow 1/2m(\Delta_0 -m)$) in the cited papers~\cite{Dvoegl-IJTP}
from the fermion-fermion case. More accurate considerations
of the fermion-boson and boson-boson interactions in the framework
of the Weinberg theory has been reported elsewhere~\cite{Dvoegl-Valladolid}.
So, the conclusion of this Section is: one can construct an analog of the Feynman-Dyson
propagator for the $2(2S+1)$ model and, hence, a `local'
theory provided that the Weinberg states are
quadrupled  ($S=1$ case).

\section{The Self/Anti-self Charge Conjugate Construct in the $(1/2,0)\oplus (0,1/2)$ Representation.}

The first formulations with doubling  solutions of the Dirac equations have been  presented in Refs.~\cite{Markov}, 
and~\cite{BarutZiino}. The group-theoretical basis for such doubling has been given
in the papers by Gelfand, Tsetlin and Sokolik~\cite{Gelfand}, who first presented 
the theory later called as `the Bargmann-Wightman-Wigner-type quantum field theory'. 
M. Markov wrote long ago {\it two} Dirac equations with  the opposite signs at the mass term~\cite{Markov}:\footnote{I turn to the pseudo-Euclidean
metric because it is more usable in the recent literature.}
\begin{eqnarray}
\left [ i\gamma^\mu \partial_\mu - m \right ]\Psi_1 (x) &=& 0\,,\\
\left [ i\gamma^\mu \partial_\mu + m \right ]\Psi_2 (x) &=& 0\,.
\end{eqnarray}
Of course, these two equations are equivalent each other on the free level since we are convinced that
the relative intrinsic parity has physical significance only.
In fact, he studied all properties of this relativistic quantum model while he did not know yet the quantum
field theory in 1937. Next, he added and  subtracted these equations. As the result the equations are
\begin{eqnarray}
i\gamma^\mu \partial_\mu \varphi (x) - m \chi (x) &=& 0\,,\\
i\gamma^\mu \partial_\mu \chi (x) - m \varphi (x) &=& 0\,.
\end{eqnarray}
Thus, $\varphi$ and $\chi$ solutions can be presented as some superpositions of the Dirac 4-spinors $u-$ and $v-$.
These equations, of course, can be identified with the equations for the Majorana-like $\lambda -$ and $\rho -$ spinors, which we presented 
in Ref.~\cite{Ahlu-NP,Dvoegl-NP}.\footnote{Of course, the signs at the mass terms
depend on, how do we associate the positive- or negative- frequency solutions with $\lambda$ and $\rho$.}
The  four-component Majorana-like spinors are defined as
	\begin{equation}
	\lambda({\bf p})= \left(\begin{array}{c}
 	\vartheta\Theta \phi^\ast_L({\bf p})\\
	\phi_L ({\bf p})
	\end{array}\right).\label{eq:taup}
	\end{equation}
They become eigenspinors of the charge conjugation operator with eigenvalues $\pm 1$ if the phase $\vartheta$ is set to  $\pm \, i$:
	\begin{equation}
		S_c\; \lambda({\bf p})\Big\vert_{\vartheta=\pm i}
                = \pm \lambda({\bf p})\Big\vert_{\vartheta=\pm i}. \label{eq:tau}
	\end{equation}
In the similar way one can construct $\rho-$ spinors on using $\phi_R$.
The dynamical equations are:
\begin{eqnarray}
i \gamma^\mu \partial_\mu \lambda^S (x) - m \rho^A (x) &=& 0 \,,
\label{11}\\
i \gamma^\mu \partial_\mu \rho^A (x) - m \lambda^S (x) &=& 0 \,,
\label{12}\\
i \gamma^\mu \partial_\mu \lambda^A (x) + m \rho^S (x) &=& 0\,,
\label{13}\\
i \gamma^\mu \partial_\mu \rho^S (x) + m \lambda^A (x) &=& 0\,.
\label{14}
\end{eqnarray}
Neither of them can be regarded as the Dirac equation.
However, they can be written in the 8-component form as follows:
\begin{eqnarray}
\left [i \Gamma^\mu \partial_\mu - m\right ] \Psi_{_{(+)}} (x) &=& 0\,,
\label{psi1}\\
\left [i \Gamma^\mu \partial_\mu + m\right ] \Psi_{_{(-)}} (x) &=& 0\,,
\label{psi2}
\end{eqnarray}
with
\begin{eqnarray}
&&\hspace{-20mm}\Psi_{(+)} (x) = \pmatrix{\rho^A (x)\cr
\lambda^S (x)\cr},
\Psi_{(-)} (x) = \pmatrix{\rho^S (x)\cr
\lambda^A (x)\cr},\,\Gamma^\mu =\pmatrix{0 & \gamma^\mu\cr
\gamma^\mu & 0\cr}
\end{eqnarray}
It is easy to find the corresponding projection operators, and the Feynman-Dyson-Stueckelberg propagator.

You may say that all this is just related to the spin-parity basis rotation (unitary transformations). 
In the previous papers the connection with the Dirac spinors has 
been found~\cite{Dvoegl-NP,Kirchbach}.
For instance,
\begin{eqnarray}
\pmatrix{\lambda^S_\uparrow ({\bf p}) \cr \lambda^S_\downarrow ({\bf p}) \cr
\lambda^A_\uparrow ({\bf p}) \cr \lambda^A_\downarrow ({\bf p})\cr} = {1\over
2} \pmatrix{1 & i & -1 & i\cr -i & 1 & -i & -1\cr 1 & -i & -1 & -i\cr i&
1& i& -1\cr} \pmatrix{u_{+1/2} ({\bf p}) \cr u_{-1/2} ({\bf p}) \cr
v_{+1/2} ({\bf p}) \cr v_{-1/2} ({\bf p})\cr},\label{connect}
\end{eqnarray}
provided that the 4-spinors have the same physical dimension.
Thus, we can see
that the two 4-spinor systems are connected by the unitary transformations, and this represents
itself the rotation of the spin-parity basis. However, it is usually assumed that the $\lambda-$ and $\rho-$ spinors describe the neutral particles,
meanwhile $u-$ and $v-$ spinors describe the charged particles. Kirchbach~\cite{Kirchbach} found the amplitudes for 
neutrinoless double beta decay ($00\nu\beta$) in this scheme. It is obvious from (\ref{connect}) that there are some additional terms comparing with the standard formulation.  

One can also re-write the above equations into the two-component forms. Thus, one obtains the Feynman-Gell-Mann  equations~\cite{FG}.
As Markov wrote himself, he was expecting ``new physics" from these equations. 

Barut and Ziino~\cite{BarutZiino} proposed yet another model. They considered
$\gamma^5$ operator as the operator of the charge conjugation. Thus, the charge-conjugated
Dirac equation has the different sign comparing with the ordinary formulation:
\begin{equation}
[i\gamma^\mu \partial_\mu + m] \Psi_{BZ}^c =0\,,
\end{equation}
and the so-defined charge conjugation applies to the whole system,  fermion + electro\-magnetic field, $e\rightarrow -e$
in the covariant derivative. The superpositions of the $\Psi_{BZ}$ and $\Psi_{BZ}^c$ give us 
the `doubled Dirac equation', as the equations for $\lambda-$ and $\rho-$ spinors. 
The concept of the doubling of the Fock space has been
developed in the Ziino works (cf.~\cite{Gelfand,D1}) in the framework of the quantum field theory. 
In their case the self/anti-self charge conjugate states
are simultaneously the eigenstates of the chirality.
It is interesting to note that for the Majorana-like field operators ($a_\eta ({\bf p}) = b_\eta ({\bf p})$) we have
\begin{eqnarray}
\lefteqn{\left [ \nu^{^{ML}} (x^\mu) + {\cal C} \nu^{^{ML\,\dagger}} (x^\mu) \right
]/2 = \int {d^3 {\bf p} \over (2\pi)^3 } {1\over 2E_p} } \\
&&\sum_\eta \left
[\pmatrix{i\Theta \phi_{_L}^{\ast \, \eta} ({\bf p}) \cr 0\cr} a_\eta
({\bf p})  e^{-ipx} +
\pmatrix{0\cr
\phi_L^\eta ({\bf p})\cr } a_\eta^\dagger ({\bf p}) e^{ipx} \right ],\\
\lefteqn{\left [\nu^{^{ML}} (x^\mu) - {\cal C} \nu^{^{ML\,\dagger}} (x^\mu) \right
]/2 = \int {d^3 {\bf p} \over (2\pi)^3 } {1\over 2E_p}} \\
&&\sum_\eta \left
[\pmatrix{0\cr \phi_{_L}^\eta ({\bf p}) \cr } a_\eta ({\bf p})  e^{-ipx}
+
\pmatrix{-i\Theta \phi_{_L}^{\ast\, \eta} ({\bf p})\cr 0
\cr } a_\eta^\dagger ({\bf p}) e^{ipx} \right ] 
\end{eqnarray}
which naturally lead to the Ziino-Barut scheme of massive chiral
fields, Ref.~\cite{BarutZiino}.

\section{The Controversy.}

I cite Ahluwalia {\it et al.}, Ref.~\cite{Ahlu-PR}:
``{\it To study the locality structure of the fields $\Lambda(x)$ and $\lambda(x)$, we observe that field momenta  are
	\begin{equation}
		\Pi(x)= \frac{\partial{\mathcal L}^\Lambda}
		{\partial\dot\Lambda} =
		\frac{\partial}{\partial t}\stackrel{\neg}{\Lambda}(x),\quad
\end{equation}
 and similarly $ \pi(x) = \frac{\partial}{\partial
 t}\stackrel{\neg}{\lambda}(x)$.  The calculational details for the
 two fields now differ significantly. We begin with the evaluation of
 the equal time anticommutator for  $\Lambda(x)$ and its conjugate
 momentum
	\begin{eqnarray} &&\{\Lambda({\bf x},t),\; \Pi({\bf x}^\prime,t)\} =
	i\int\frac{d^3 p}{(2\pi)^3}\frac{1}{2 m} e^{i {\mathbf p}\cdot
	({\mathbf x}-{\mathbf x}^\prime)} \nonumber\\
	&&\times\underbrace{\sum_\alpha\left[ \xi_\alpha({\bf p})
	\stackrel{\neg}{\xi}_\alpha({\bf p}) - \zeta_\alpha(- {\bf p})
	\stackrel{\neg}{\zeta}_\alpha(- {\bf p})\right]}_{=\, 2 m [ I +
	{\mathcal G}(\mathbf{p})]} .\nonumber 
	\end{eqnarray}
The term containing
 ${\mathcal G}({\bf p})$ vanishes only when ${\mathbf x}-{\mathbf x}^\prime$ lies along the $z_e$ axis (see Eq.~(24) [therein], and discussion of this integral in Ref.~\cite{AhluwaliaGRU})
	\begin{equation}
		{\mathbf x}-{\mathbf x}^\prime \;\mbox{along }z_e:\quad \{\Lambda({\bf x},t),\; \Pi({\bf x}^\prime,t)\} 
		=  i \delta^3({\bf x} -{\bf x}^\prime) I
\label{eq:LPac}
	\end{equation}
The anticommutators for the particle/antiparticle annihilation and creation
operators suffice to yield the remaining locality conditions,
	\begin{equation}
		\{\Lambda({\bf x},t),\; \Lambda({\bf x}^\prime,t)\} =
O
,\quad \{\Pi({\bf x},t),\; \Pi({\bf x}^\prime,t)\}
		\label{eq:LLPPac} = 
O
.
	\end{equation}
The set of anticommutators contained in Eqs.~(\ref{eq:LPac}) and
(\ref{eq:LLPPac}) establish that $\Lambda(x)$ becomes local along the $z_e$ axis. For this reason we call $z_e$ as the dark axis of locality.}"

Next, I cite Rodrigues {\it et al.}, Ref.~\cite{Rodrigues-PR}:
``{\it We have shown through explicitly and detailed calculation that the integral of
$\mathcal{G}(\mathbf{p})$ appearing in Eq.(42) of \cite{Ahlu-PR} \ is null for
$\mathbf{x-x}^{\prime}$ lying in three orthonormal spatial directions in the
rest frame of an arbitrary inertial frame $\mathbf{e}_{0}=\partial/\partial t$.

This shows that the existence of elko spinor fields does not implies in any
breakdown of locality concerning the anticommutator of $\{\Lambda
(\mathbf{x,}t),\Pi(\mathbf{x}^{\prime},t\}$ and moreover does not implies in
any preferred spacelike direction field in Minkowski spacetime.}"

Who is right? In 2013 W. Rodrigues~\cite{Rodrigues-IJTP} changed a bit his opinion. He wrote: 
``{\it When $\Delta_{z}\neq0$, $\mathcal{\hat{G}}(\mathbf{x-x}^{\prime})$ is null the
anticommutator is \emph{local} and thus there exists in the elko theory\ as
constructed in \cite{Ahlu-PR} an infinity number of
\emph{\textquotedblleft}locality\ directions\emph{\textquotedblright.} On the
other hand $\mathcal{\hat{G}}(\mathbf{x-x}^{\prime})$ is\ a distribution with
support in $\Delta_{z}=0$. So$,$ the directions $\mathbf{\Delta}=(\Delta
_{x},\Delta_{y},0)$\ are nonlocal in each arbitrary inertial reference frame
$\mathbf{e}_{0}$ chosen to evaluate $\mathcal{\hat{G}}(\mathbf{x-x}^{\prime}%
)$}", thus accepting the Ahluwalia {\it et al.} viewpoint. See the cited papers for the notation.

Meanwhile, I suggest to use the 8-component (or 16-component) formalism (see the Section 2) in the similarity with the 12-component formalism of the Section 1. If we calculate
\begin{eqnarray}
\lefteqn{\hspace{-6mm}S_F^{(+,-)} (x_2, x_1) =\int \frac{d^3 p}{(2\pi)^3}\frac{m}{E_p} \left [
\theta (t_2 -t_1) \, a \,\,
\Psi_\pm^\sigma (p)  \overline \Psi_\pm^{\sigma} (p) e^{-ipx} + \right. }\nonumber\\
&&\left. \qquad\qquad  +\, \,\theta (t_1 - t_2) \, b \,\,
\Psi_\mp^\sigma (p) \overline \Psi_\mp^\sigma (p) e^{ipx} \right ]= \nonumber\\
&=&\int 
\frac{d^4 p}{(2\pi)^4} e^{-ipx} \frac{(\hat p\pm m)}{p^2 -m^2
+i\epsilon}, 
\end{eqnarray}
we easily come to the result that the corresponding Feynman-Dyson propagator gives  the local theory 
in the sense:
\begin{equation}
\sum_{\pm} [ i\Gamma_\mu \partial^\mu_2 \mp m] S_F^{(+,-)} (x_2 - x_1) = \delta^{(4)} (x_2 - x_1).
\end{equation}
However, physics should choose only one correct formalism. It is not clear, why two correct mathematical formalisms  lead to different physical results? First of all, we should check, whether this possible non-locality in the propagators has influence on the physical observables such as the scattering amplitudes, the energy spectra and the decay widths. If not, we may find some unexpected symmetries in relativistic quantum mechanics/field theory. This is the task for future publications.

{\bf Note Added.} The dilemma of the (non)local propagators for the spin $S=1$ has also been analized in~\cite{Kruglov2}
within the Duffin-Kemmer-Petiau (DKP) formalism or the Dirac-K\"ahler formalism~\cite{Kruglov1}. That author mentions se\-ve\-ral works which 
``showed that [the Dirac-K\"ahler equation] is equivalent to four Dirac equations" for spin-1/2 particles. However, the field, which Kruglov 
associated with the Dirac-K\"ahler equation, transforms according to
the  $(1,0)+(0,1) + 2 (0,0) + 2 (1/2,1/2)$ Lorentz group representation.
The Dirac field function transforms according to the $(1/2,0)+(0,1/2)$ representation. In the case of 16 components 
we should have $4 [(1/2,0)+(0,1/2)]$. Even if we  combine four $\gamma$-matrices, the new $16\times 16$ matrix should satisfy the same 
anticommutation relations (the definition of the gamma matrices, indeed) as the $4\times 4$ $\gamma$-matrix. It also should transform
accordingly:
\begin{equation}
S (\Lambda) \gamma^\mu_{16\times 16} S^{-1} (\Lambda) = (I_{4\times 4}\otimes\Lambda^{-1})^\mu_\nu \gamma^\nu_{16\times 16}\,,
\end{equation}
$\otimes$ denoted the Kronecker product.
It is also possible to note that the $(1,0)+(0,0)$ (or $(0,1)+(0,0)$) field can be put in the 4-component form. The vector of the $(1/2,1/2)$ 
representation has 4 components. Then, it is possible to form four 4-component equations, and write them in the $16\times 16$ matrix form. 
However, the $4\times 4$ $\alpha$-matrices are not the $\gamma$-matrices, 
as shown by Dowker~\cite{Dowker} long ago (cf.~\cite{DVO-NCB}). Next, let us investigate the transformation properties.
We  first consider the four Dirac equations. So, we have
\begin{equation}
S (\Lambda) = I_{16\times 16} - {i\over 4} [I_{4\times 4}\otimes \sigma_{\mu\nu}]\omega^{\mu\nu}+\ldots
\end{equation}
In the case of the DKP representation we have different transformation properties of the $16\times 16$ field function:
the entries of the antisymmetric tensor field, 4-vectors, scalars represent themselves the Kronecker products of the dotted and undotted 
spinors, see~\cite{Landau}. The $S(\Lambda)$ matrix is now
\begin{eqnarray}
S (\Lambda) &=& (I_{4\times 4} - {i\over 4}\sigma_{\mu\nu}\omega^{\mu\nu}+\ldots )\otimes 
(I_{4\times 4} - {i\over 4}\sigma_{\alpha\beta}\omega^{\alpha\beta}+ \ldots) = \nonumber\\
&=&I_{16\times 16} - {i\over 4} [ I_{4\times 4}\otimes \sigma_{\mu\nu}
+ \sigma_{\mu\nu}\otimes I_{4\times 4}] \omega^{\mu\nu}+\ldots
\end{eqnarray}
The Kronecker product is non-commutative. For example,for ${\vec\alpha} = diag ({\vec \sigma}\quad -{\vec \sigma})$ we have:
\begin{eqnarray}
&&{\vec \alpha}^1\otimes I_{4\times 4} = \pmatrix{0&I&0&0\cr
I&0&0&0\cr
0&0&0&-I\cr
0&0&-I&0\cr},\,
{\vec \alpha}^2\otimes I_{4\times 4} = \pmatrix{0&-iI&0&0\cr
iI&0&0&0\cr
0&0&0&iI\cr
0&0&-iI&0\cr},\nonumber\\
&&{\vec \alpha}^3\otimes I_{4\times 4} = \pmatrix{I&0&0&0\cr
0&-I&0&0\cr
0&0&-I&0\cr
0&0&0&I\cr}\,.
\end{eqnarray}
On the other hand,
\begin{equation}
I_{4\times 4} \otimes {\vec \alpha} = \pmatrix{{\vec \alpha}&0&0&0\cr
0&{\vec \alpha}&0&0\cr
0&0&{\vec \alpha}&0\cr
0&0&0&{\vec \alpha}\cr
}\,.
\end{equation}
So, the transformation rules will not be similar even for double arguments (parameters).
The mapping of the DKP equations onto the four Dirac equations is doubtful even if the corresponding $16\times 16$ matrix 
belongs to the similar Clifford Algebra.

Next, the propagators given in~\cite{Kruglov2} are those in the generalized Duffin-Kemmer-Petiau formalism, in fact. They are 
not in the Weinberg-Tucker-Hammer  formalism. Meanwhile, the connection
between the components of the DKP formalism and those of the 2(2S+1) formalism is known since long ago 
(for example, see the Greiner book on relativistic
quantum mechanics~\cite{Greiner}). However, it is doubtful that
any physically plausible relations exist between mentioned propagators. This is again due to different Lorentz group representations,
the $(1,0)+(0,1)$ (or $2[(1,0)+(0,1)]$, or $4[(1,0)+(0,1)]$) representations commented in this paper,
and the $(1,0)+(0,1)+ 2(0,0) +2(1/2,1/2)$ representation of the DKP formalism. Moreover, the problem of the massless limit
was not discussed in the DKP formalism, which is non-trivial (like that of the Proca formalism~\cite{Stepan}).

\bigskip

{\bf Acknowledgments.} I acknowledge discussions with Prof. W. Rodrigues, Jr. and Prof. Z. Oziewicz.
I am grateful to the Zacatecas University for professorship.

}

\end{document}